\documentclass[aps,pra,twocolumn,superscriptaddress]{revtex4-2}
\usepackage{mathtools}
\usepackage{graphicx}
\usepackage{hyperref}
\usepackage{enumerate}

\newcommand{\abs}[1]{| #1 |}

\begin{document}
\title{Direct measurement of coherent light proportion from a practical laser source}
\author{Xi~Jie~Yeo}
\affiliation{Centre for Quantum Technologies, National University of Singapore, 3 Science Drive 2, Singapore 117543}

\author{Eva~Ernst}
\affiliation{Centre for Quantum Technologies, National University of Singapore, 3 Science Drive 2, Singapore 117543}
% \affiliation{Institute fuer Experimentalphysik, Universitaet Innsbruck, Technikerstrasse 25, 6020 Innsbruck, Austria}

\author{Alvin~Leow}
\affiliation{Department of Physics, National University of Singapore, 2 Science Drive 3, Singapore 117551}

\author{Jaesuk~Hwang}
\affiliation{Centre for Quantum Technologies, National University of Singapore, 3 Science Drive 2, Singapore 117543}

\author{Lijiong~Shen}
\affiliation{Centre for Quantum Technologies, National University of Singapore, 3 Science Drive 2, Singapore 117543}

\author{Christian~Kurtsiefer}
\affiliation{Centre for Quantum Technologies, National University of Singapore, 3 Science Drive 2, Singapore 117543}
\affiliation{Department of Physics, National University of Singapore, 2 Science Drive 3, Singapore 117551}
\email[]{christian.kurtsiefer@gmail.com}

\author{Peng~Kian~Tan}
\affiliation{Centre for Quantum Technologies, National University of Singapore, 3 Science Drive 2, Singapore 117543}

\date{\today}
\begin{abstract}
We present a technique to estimate the proportion of coherent emission in the
light emitted by a practical laser source without spectral filtering. The
technique is based on measuring interferometric photon correlations between
the output ports of an asymmetric Mach-Zehnder interferometer.
With this, we characterize the fraction of coherent emission in the light
emitted by a laser diode when transiting through the lasing threshold. 
\end{abstract}

% insert suggested PACS numbers in braces on next line
% \pacs{}

\maketitle
\section{Introduction}
The invention of lasers can be traced to work describing the emission process
of the light from an atom to be spontaneous or stimulated
\cite{Einstein1916}. An ensemble of atoms undergoing stimulated emission will
emit coherent light that has a well-defined phase, while 
spontaneous emission will lead to randomly phased incoherent light
\cite{Loudon2000}. Coherent light is at the core of many applications, including
interferometry \cite{Hariharan2007}, metrology \cite{Wright1969}, and optical
communication.

In traditional models of macroscopic lasers \cite{Lamb1964,Arecchi1965,Haken1975}, 
the emitted light is modeled to originate dominantly from stimulated emission. 
These models predict a phase transition of the nature of emission with
increasing pump strength, separating two regimes where light emitted is either
spontaneous (below threshold), or stimulated (above threshold). 

However, experiments on small lasers have shown that the transition from spontaneous to stimulated emission is not abrupt
\cite{Choi2007,Urlich2007,Wiersig2009,Hostein2010,Kreinberg2017}.
Instead, light emitted from the laser can be described as
a mixture of spontaneous and stimulated emission across a transition range.

In these experiments, the transition from spontaneous to stimulated emission was characterized by measuring
the second-order photon correlation $g^{(2)}$,
using a Hanbury-Brown and Twiss scheme \cite{HBT1956}.
The measurement result can be explained using Glauber's theory of optical coherence \cite{Glauber1963},
where incoherent light from spontaneous emission would exhibit a ``bunching''
signature with $g^{(2)}(0)>1$,
while coherent light from stimulated emission exhibits a Poissonian
distribution with $g^{(2)}=1$.

The ``bunching'' signature associated with incoherent light has a
characteristic timescale inversely proportional to its spectral width
according to the Wiener-Khintchine theorem~\cite{Wiener1930,Khintchine1934,MandelWolf1995}. 
In a practical measurement, the amplitude of the ``bunching'' signature scales
with the ratio of characteristic timescale of the  light
to the timing response of the detectors \cite{Scarl1968}.
Thus, when the spectral width of incoherent light is so broad
that the characteristic timescale of the ``bunching'' signature is smaller than the detector timing uncertainty,
incoherent light may exhibit $g^{(2)}\approx1$, 
like coherent light.

To overcome the limited detector timing uncertainty, 
a narrow band of incoherent light can be prepared with filters from a wide
optical spectrum of an incoherent licht source~\cite{PK2017}.
The narrow spectral width of a filtered incoherent light has a correspondingly
larger characteristic coherence timescale,
which may be long enough to be resolvable by the detectors.

However, for characterising the transition of a laser from spontaneous to
stimulated emission, such spectral filtering presents some shortcomings.
First, as spectral filtering discards light outside the transmission window of
a filter, a result would be inconclusive for the full emission of the source.
Second,
spectral filtering requires \textit{a priori} information or an educated guess of the central frequency and bandwidth of stimulated emission.
Third, it has been shown that spectral filtering below the Schawlow-Townes linewidth of the laser results in $g^{(2)}(0)>1$, similar to light from spontaneous emission \cite{Neelen1992}.

Light emitted by a laser is also incoherent in multimode operation
\cite{Weber1970,Yin1996}, where a laser may emit coherent light in multiple
transverse 
and/or longitudinal modes. The light in each mode may be coherent, but a
combination of multiple modes may result in a randomly phased light, and
therefore appear incoherent. 

This motivates for methods quantifying the proportion of coherent light
emitted by a source without the need for spectral filtering.
A method to characterise the stimulated and spontaneous emission from a pulsed
laser has been demonstrated before \cite{George2021}.

In this paper, we present a method to quantify bounds for the proportion of
coherent light for a continuous wave laser. Specifically, we investigate the
brightest mode of coherent emission from a semiconductor laser diode
by using interferometric photon correlations, i.e., 
a correlation of photoevents detected at the output ports 
of an asymmetric Mach-Zehnder interferometer.
Earlier methods of interferometric photon correlation measurements were used to study
spectral diffusion in organic molecules embedded in solid matrix~\cite{brokmann2006photon,PhysRevA.76.033824}.
The method of interferometric photon correlation we use here was
originally applied to differentiate between incoherent light and coherent light with amplitude fluctuations~\cite{lebreton2013exp}.
In contrast to second-order photon correlations, 
this method can clearly distinguish between finite linewidth coherent light and broadband incoherent light~\cite{Lebreton2013Theory}.
We use this method to extract the fraction of coherent light emitted by the laser diode
over a range of pump powers across the lasing threshold.

%%%%%%%%%%%%%%%%%%%%%%%%%%%%%%%%%%%%%%%%%%%%%%%
\section{Interferometric Photon Correlations}
%%%%%%%%%%%%%%%%%%%%%%%%%%%%%%%%%%%%%%%%%%%%%%%

\begin{figure}
  \begin{center}
    \includegraphics[width=0.8\columnwidth]{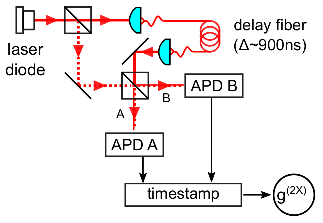}
  \end{center}
  \caption{Experimental setup for measuring interferometric photon
    correlations.
    Light from a laser diode enters an asymmetric Mach-Zehnder Interferometer.
    Single photon avalanche photodetectors (APD) 
    at each output port of the interferometer
    generate photodetection events, which are time-stamped to
    extract the correlations numerically.} \label{fig:setup}
\end{figure}

The setup for an interferometric photon correlation measurement $g^{(2X)}$ is shown in Fig.~\ref{fig:setup}.
Light emitted by the laser diode
is sent through an asymmetric Mach-Zehnder interferometer, with a 
propagation delay $\Delta$ between the two paths of the interferometer that
exceeds the coherence time of the light.

With a light field $E(t)$ at the input,
the light fields at the output ports $A,B$ of the interferometer are 
\begin{equation}
    E_{A,B}(t) = \frac{E(t) \pm E(t+\Delta)}{\sqrt{2}}\,,
    \label{eqn:fields}
\end{equation}
with the relative phase shift $\pi$ acquired by one of the output fields from
the beamsplitter. 

Using these expressions for the electrical fields, 
the temporal correlation of
photodetection events between the two output ports is given by
\begin{equation}
  g^{(2X)}(t_{2}-t_{1})=
   \frac{\langle E^{*}_{A}(t_{1})E^{*}_{B}(t_{2})E_{B}(t_{2})E_{A}(t_{1})\rangle}{ \langle E^{*}_{A}(t_{1})E_{A}(t_{1})\rangle \langle E^{*}_{B}(t_{2})E_{B}(t_{2})\rangle}\,.
\end{equation}

Therein, $\langle \rangle$ indicates an expectation value.
 and/or an ensemble average.
Using Eqn.~\ref{eqn:fields}, 
$g^{(2X)}(t_{2}-t_{1})$ can be grouped in several terms: 
\begin{equation}
  \begin{aligned}
   g^{(2X)}&(t_{2}-t_{1})=\\
          = \frac{1}{4}[ &\langle {E^{*}(t_{1})E^{*}(t_{2})E(t_{2})E(t_{1})} \rangle \\
        & + \langle {E^{*}(t_{1}+\Delta)E^{*}(t_{2}+\Delta)E(t_{2}+\Delta)E(t_{1}+\Delta)} \rangle \\[1mm]
        & + \langle {E^{*}(t_{1}+\Delta)E^{*}(t_{2})E(t_{2})E(t_{1}+\Delta)} \rangle\\
        & + \langle {E^{*}(t_{1})E^{*}(t_{2}+\Delta)E(t_{2}+\Delta)E(t_{1})} \rangle\\[1mm]
        & - \langle {E^{*}(t_{1}+\Delta)E^{*}(t_{2})E(t_{2}+\Delta)E(t_{1})} \rangle\\
        & - \langle {E^{*}(t_{1})E^{*}(t_{2}+\Delta)E(t_{2})E(t_{1}+\Delta)} \rangle]\,.\\
  \end{aligned}
  \label{eqn:expansion}
\end{equation}

The first two terms have the form of conventional second-order photon
correlation functions $g^{(2)}(t_2-t_1)$.
The next two terms are conventional second-order photon correlation functions,
time-shifted forward and backward in their argument by the propagation delay
$\Delta$. The last two terms reduce $g^{(2X)}$, leading to a dip
at zero time difference $t_2-t_1 = 0$,
with a width given by the coherence time of the light.

The expectation values appearing in Eqn.~\ref{eqn:expansion} %for $g^{(2X)}$
can be evaluated by using 
statistical expressions~\cite{Loudon2000} of $E(t)$ for incoherent and coherent light \cite{Lebreton2013Theory}.

For incoherent light, 
$g^{(2X)}$ exhibits a ``bunching'' signature peaking at time differences
$\pm\Delta$, $g^{(2X)}(\pm\Delta) = 1+(1/4)$.
At zero time difference, 
the expected ``bunching'' signature from conventional second-order photon
correlation functions in the first two terms,
and the dip from the last two terms of Eqn.~\ref{eqn:expansion} cancel each
other, resulting in $g^{(2X)}(0)= 1$.

For coherent light, 
the second-order photon correlation function $g^{(2)}=1$ combines with the
negative contributions from the last two terms of Eqn.~\ref{eqn:expansion}
such that $g^{(2X)}(0) = 1/2$. % (see Fig.~\ref{fig:sample_plots} bottom).

%%%%%%%%%%%%%%%%%%%%%%%%%%%%%%%%%%%%%%%%%%%%%%%%
\section{Fraction of coherent light in a mixture}
%%%%%%%%%%%%%%%%%%%%%%%%%%%%%%%%%%%%%%%%%%%%%%%%
In order to obtain an interpretation of the nature of the light emitted
beyond just presenting the components of $g^{(2X)}$, we consider a light field
that is neither completely coherent nor incoherent.
We assume that light emitted by the laser is a mixture of coherent
light field $E_\text{coh}$, and a light field $E_\text{unc}$ uncorrelated to $E_\text{coh}$.
The nature of $E_\text{unc}$ can be coherent, incoherent, or a coherent-incoherent mixture.
In the following, we extract quantitative information about the
components of the light field from interferometric photon correlations
$g^{(2X)}$, namely the fraction of optical power in the brightest coherent
component. 

We model the light field mixture with an electrical field
\begin{equation}
    E_\text{mix}(t) = \sqrt{\rho}E_\text{coh}(t) + \sqrt{1-\rho}E_\text{unc}(t)\,,
    \label{eqn:mixture_field}
\end{equation}
where $\rho$ is the
fraction of optical power the brightest coherent emission, and
the respective light field terms are normalised 
such that $\abs{E_\text{mix}} = \abs{E_\text{coh}} = \abs{E_\text{unc}}$.

Evaluating photon correlation in Eqn.~\ref{eqn:expansion} with this light model,
and further assuming that first, the propagation delay in the interferometer is
significantly longer than the coherence time scale of the light source,
and second, the interferometer has good visibility yields
\begin{equation}
    g_{\text{mix}}^{(2X)}(0) = 2\rho-\frac{3\rho^{2}}{2} + \frac{(1-\rho)^{2}}{2}g_{\text{unc}}^{(2)}(0)
 \label{eqn:g2x_mixture_extractable}
\end{equation}
at zero time difference, with only two remaining parameters, $\rho$ 
and $g_{\text{unc}}^{(2)}(0)$, 
the second order photon correlation of the uncorrelated
field at zero time difference (see Appendix~\ref{app:g2x_mixture}). 

The connection in Eqn.~\ref{eqn:g2x_mixture_extractable}, together with the
physical requirement $0 \leq \rho \leq 1$ for the fraction of coherent light
limits the possible combinations of  $g_{\text{unc}}^{(2)}(0)$ and
$g_{\text{mix}}^{(2X)}(0)$, shown as non-shaded areas in
Fig.~\ref{fig:bounds_plots}; the exact expressions for the boundaries are given
in Appendix~\ref{app:boundaries}.

\begin{figure}
    \begin{center}
    \includegraphics[width=\linewidth]{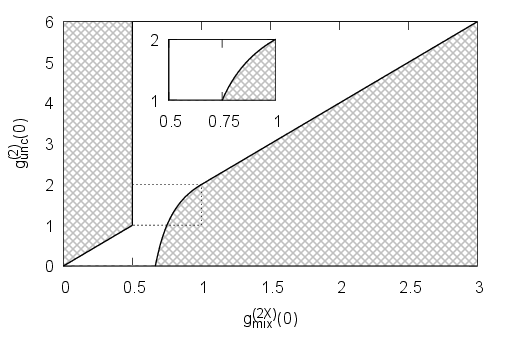}    
    \end{center}
  \caption{Combinations of $g_{\text{unc}}^{(2)}(0)$ and
    $g_{\text{mix}}^{(2X)}(0)$ that correspond to physical and real-valued
    $\rho$. In shaded areas, no such solution exist. Inset: Zoom into the region $1\leq g_{\text{unc}}^{(2)}(0)\leq 2$, where the uncorrelated light source is assumed to be a mixture of coherent and completely incoherent light, and thermal light.} 
  \label{fig:bounds_plots}
\end{figure}

We can now further assume that the uncorrelated light source generates some
mixture of coherent and completely incoherent light ($g^{(2)}(0)=1$),
and thermal light ($g^{(2)}(0)=2 $). This constrains the second-order
photon correlation of the uncorrelated light:
\begin{equation}
    1\leq g_{\text{unc}}^{(2)}(0)\leq 2.
    \label{eqn:g2_bounds}
\end{equation}
We impose these bounds in Eqn.~\ref{eqn:g2x_mixture_extractable},
and extract the bounds to the fraction of optical power in the brightest coherent emission $\rho$   
with an upper bound,
\begin{equation}
     \rho \leq \sqrt{2 - 2\,g^{(2X)}(0)},
 \label{eqn:fraction_upper_bound}
\end{equation}
and a lower bound,
\begin{equation}
    \rho \geq  \begin{cases}\frac{1}{2}+\frac{1}{2}\sqrt{3-4\,g^{(2X)}(0)}, &\text{for}\, \frac{1}{2} \leq g^{(2X)}(0) \leq \frac{3}{4} \\
    2-2\,g^{(2X)}_{\text{mix}}(0), &\text{for}\, \frac{3}{4} \leq g^{(2X)}(0) \leq 1\end{cases}\,,
 \label{eqn:fraction_lower_bounds}
\end{equation}
with $g^{(2X)}_{\text{mix}}(0)$ ranging from $1/2$ for fully coherent light, 
to 1 for fully incoherent light.

In practice, these two bounds for $\rho$ are
quite tight, and allow to extract the fraction $\rho$ in an
experiment with a small uncertainty.

%%%%%%%%%%%%%%%%%%%%%%%%%%%%%%%%%%%%%%%%%%%%%%%%
\section{Experiment}
%%%%%%%%%%%%%%%%%%%%%%%%%%%%%%%%%%%%%%%%%%%%%%%%
In our experiment,
we measure interferometric photon correlations of light 
emitted from a temperature-stabilised distributed feedback laser diode 
with a central wavelength around 780\,nm.

The setup is shown in Fig.~\ref{fig:setup}. Interferometric photon
correlations are obtained from an asymmetric Mach-Zehnder interferometer, formed by 50:50 fibre beamsplitters
and a propagation delay $\Delta$ of about 900\,ns  through a 180\,m long
single mode optical fibre in one of the arms.
Photoevents at each output port of the interferometer were detected with
actively quenched silicon single photon avalanche photo diodes (APD). 
The detected photoevents were time-stamped with a
resolution of 2\,ns for an integration time $T$.

\begin{figure}
    \begin{center}
      \includegraphics[width=\linewidth]{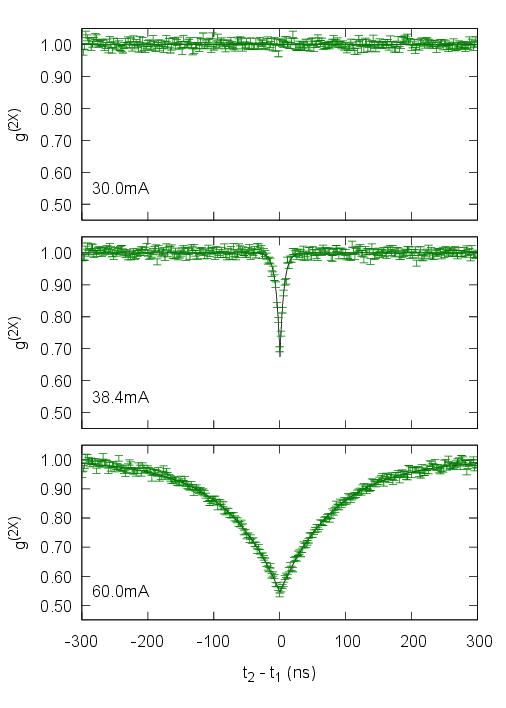} 
    \end{center}
  \caption{Interferometric photon correlations $g^{(2X)}$ for different
    laser currents $I_L$, extracted from a histogram of photodetector time
    differences (green symbols). The error range at a specific time bin
    indicates an expected uncertainty according to a Poissonian counting
    statistics. The black solid lines show a fit to  Eqn.~\ref{eqn:fit},
    resulting in values for $A$ (from top to bottom) of
    $-0.0006\pm0.0003$, $0.326\pm0.008$, $0.455\pm0.002$, respectively. } 
  \label{fig:sample_plots}
\end{figure}

The correlation function
$g^{(2X)}$ is extracted through histogramming all time differences $t_2-t_1$
between detection event pairs in the inverval $T$ numerically, 
which allows for a clean normalization. 
The resulting correlation is 
fitted to a two-sided exponential function, 
\begin{equation}
 g^{(2X)}(t_2 - t_1) = 1 -  A\cdot\exp\left(-\frac{\abs{t_2 - t_1}}{\tau_c}\right)\,,
 \label{eqn:fit}
\end{equation} 
where $\tau_c$ is the characteristic time constant of the coherent light,
and $A$ is the amplitude of the dip.
The value of $g^{(2X)}(0)$ is the extracted from the fit as $1-A$.
Examples of measured correlation functions and corresponding fits for
different laser powers are shown in Fig.~\ref{fig:sample_plots}.

%%%%%%%%%%%%%%%%%%%%
\subsection{Tranisition from incoherent to coherent light}
\begin{figure}
    \begin{center}
      \includegraphics[width=\linewidth]{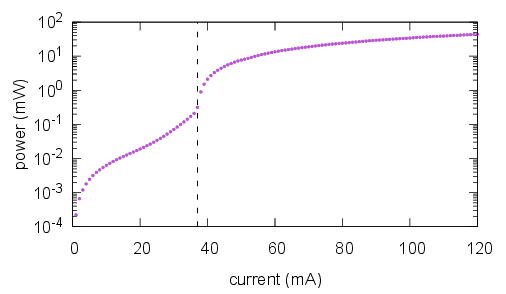} 
    \end{center}
  \caption{Measured laser power against laser current $I_L$. The sharpest change  was measured at $I_T=37$\,mA, indicating the threshold current (dashed
    line). 
  } 
  \label{fig:power_vs_current}
\end{figure}

%%%%%%%%%%%%%%%%%%%%
A transition from incoherent to coherent emission
is expected as the laser current is increased 
across the lasing threshold of the laser.
We identify the lasing threshold of a laser diode $I_T$, 
by measuring the steepest increase of optical power
with the laser current (Fig~\ref{fig:power_vs_current}). 
For our diode, we find $I_T=37$\,mA. 

\begin{figure}
    \begin{center}
      \includegraphics[width=\linewidth]{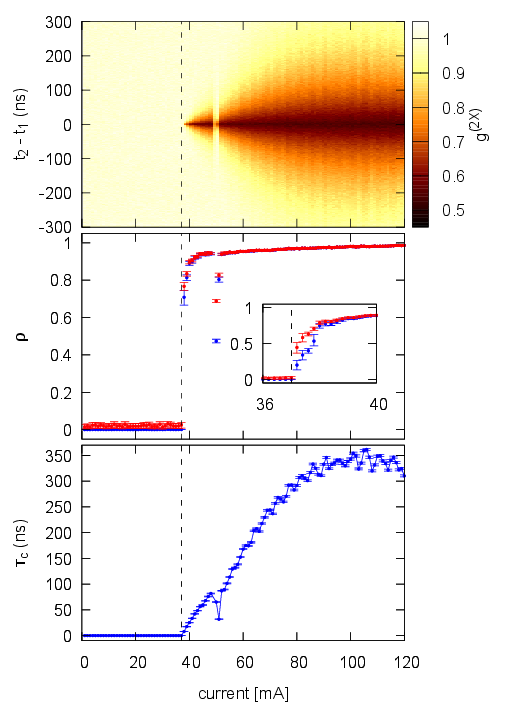}
    \end{center}
  \caption{Top:
      Interferometric photon corrrelations $g^{(2X)}$ for different laser
      currents $I_L$. 
      Middle: Corresponding upper bound of
      fraction $\rho$ of coherent light (red) extracted via 
      Eqn.~\ref{eqn:fraction_upper_bound_A},
      and the lower bound (blue) extracted via Eqn.~\ref{eqn:fraction_lower_bounds_A}
      from $g^{(2X)}(0)$.
      The dip in $\rho$ is a result of emission at multiple chip modes as explained in Section~\ref{subs:mixture}. 
      The inset shows the extracted bounds for $\rho$ at finer steps of laser current near the lasing threshold.
      Bottom: Coherence time of coherent light $\tau_{c}$ extracted from $g^{(2X)}$. 
      The dashed line indicates the threshold current $I_T=37$\,mA.
  } 
  \label{fig:fraction_vs_current}
\end{figure}

To observe the transition from incoherent to coherent emisssion,
we extract the fraction $\rho$ of optical power in the brightest coherent
component in the light field  at different laser current $I_{L}$
across the lasing threshold from measurements of $g^{(2X)}$
(Fig.~\ref{fig:fraction_vs_current}, top part). 
The amplitude of the dip is extracted by fitting these
correlations to Eqn.~\ref{eqn:fit},
from which the upper bound and lower bound of 
$\rho$ 
is extracted (Fig.~\ref{fig:fraction_vs_current}, middle part).

From the fit, $\rho$ remains near 0 below threshold.
Above the threshold $\rho$ increases quickly with $I_L$
in a phase-transition manner,
reaching $\rho=0.986$ ($90\%$ confidence interval: $0.982$ to $0.989$) at
$I_L=120$\,mA.
This agrees with the expectation that the emission of the laser diode is 
increasingly dominated by stimulated emission when driven with current above
the lasing threshold\cite{Haug1967, Haken1981}.

The upper and lower bounds for $\rho$ from Eqn.~\ref{eqn:fraction_upper_bound}
and \ref{eqn:fraction_lower_bounds} are quite tight even near the lasing
threshold, suggesting that the mixture model Eqn.~\ref{eqn:mixture_field}
captures the nature of the light through the phase transition well.

The coherence time of the coherent light $\tau_{c}$ can also be extracted
from fitting $g^{(2X)}$ measurements
to Eqn.~\ref{eqn:fit} (bottom Fig.~\ref{fig:fraction_vs_current}). 
We observe that the coherence time increases 
with the current after the threshold current, 
before reaching a steady value between 300 to 350\,ns.
The increase of coherence time corresponds to a narrowing of the emission
linewidth. This observation agrees with predictions from laser theory
that line narrowing is expected with increased pumping~\cite{Haken1981}.
A small modulation of the coherence time becomes visble for larger laser
currents, with a periodicity of about 6\,mA.

%%%%%%%%%%%%
\subsection{Light statistics near a mode hop}\label{subs:mixture}
%%%%%%%%%%%%%
Above the threshold, the laser can oscillate at different longitudinal
modes for different laser currents.
It is interesting to observe the presented method for extracting the fraction of
coherent emission near such a mode hop, where two coherent emission modes
compete.

For this, the spectrum of light emitted by the laser diode was recorded 
at different laser currents with an optical spectrum analyser
with a spectral resolution of 2\,GHz (Bristol 771B-NIR).
The laser diode emitted light into two distinct narrow spectral bands with a
changing power ratio in the laser current range between 49\,mA and 52\,mA.
Outside this window, only one of the modes could be identified.
Below 49\,mA, the laser emission was centered around 780.07\,nm, above
52\,mA around 780.34\,nm.

\begin{figure}
    \begin{center}
      \includegraphics[width=\linewidth]{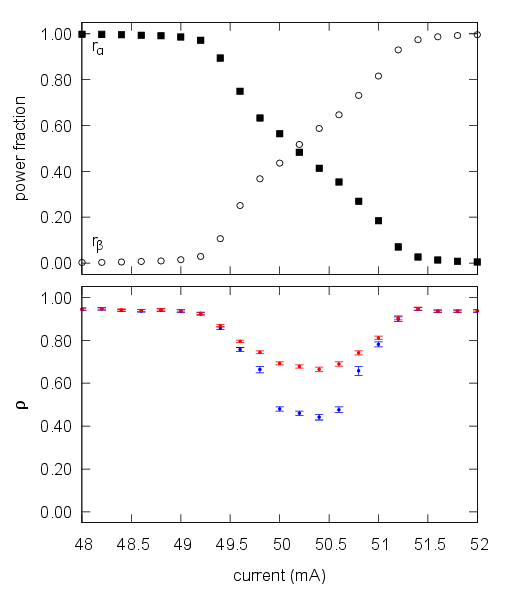}
    \end{center}
  \caption{Different chip modes of the laser diode are excited for different
    currents, resulting in a reduction of the $g^{(2X)}$ signature in a mode
    competition regime. Top: Power ratios $r_{\alpha,\beta}$ as a function of
    current for the chip modes $\alpha$ and $\beta$ % emitting in narrow bands
    around 780.07\,nm (solid squares) and 780.34\,nm (hollow circles), respectively.  Bottom:
    Upper bound of fraction $\rho$ of coherent light (red) from extracted via 
    Eqn.~\ref{eqn:fraction_upper_bound_A},
    and the lower bound (blue) extracted via Eqn.~\ref{eqn:fraction_lower_bounds_A}
    from $g^{(2X)}(0)$.
  } \label{fig:two_mode_vs_current}
\end{figure}

The power fractions $r_{\alpha, \beta}$ of these two chip modes $\alpha$ and
$\beta$ near the mode hop,
\begin{equation}
  r_{\alpha,\beta}=\frac{P_{\alpha,\beta}}{P_{\alpha} + P_{\beta}}\,,
 \label{eqn:ratio}
\end{equation} 
undergo a nearly linear transition 
(Fig.~\ref{fig:two_mode_vs_current}, top traces).

We measured $g^{(2X)}$ in the same transition regime and extract 
$\rho$ as described above (Fig.~\ref{fig:two_mode_vs_current}, bottom trace). 
In the transition regime, $\rho$ decreases when both chip modes are present.
This can be interpreted as coherent light in one emission band 
being uncorrelated to coherent light in the other one, but we did not carry
out a measurement that would test for a phase relationship between the two
modes.

%%%%%%%%%%%%%%%%%%%%
\section{Conclusion}
%%%%%%%%%%%%%%%%%%%%

We presented a method to extract the fraction of coherent light in the
emission of a laser by using interferometric photon correlations.
As a demonstration, we analyzed light emitted from a diode laser over a range
of laser currents,
and observe a continuously increasing fraction of coherent light with
increasing laser current above the lasing threshold.
Applying this technique to light emitted near a mode hop between longitudinal
modes suggests a reduction of the fraction of coherent light in the transition
regime, and an interpretation that the two longitudinal modes can be
viewed as mutually incoherent coherent emissions.
Apart from the characterisation of lasers, this method 
can be useful in practical applications of continuous-variable quantum key distribution protocols,
% ~\cite{Lodewyck2005,Weedbrook2012}
where the noise of lasers as a source of coherent states
needs to be carefully characterised to ensure security claims~\cite{Shen2009,Usenko2010,Shen2011}.

% APPPENDIX

\appendix
%%%%%%%%%%%%%%%%%%%%%%%%%%%%%%%%%%%%
\section{Interferometric photon-correlation for a mixture of light fields}\label{app:g2x_mixture}
%%%%%%%%%%%%%%%%%%%%%%%%%%%%%%%%%%%

The evaluation of $g^{(2X)}$ via Eqn.~\ref{eqn:expansion} 
requires the conventional second-order photon correlation function $g^{(2)}(t_1 -t_2)=
\langle {E^{*}(t_{1})E^{*}(t_{2})E(t_{2})E(t_{1})} \rangle$.
For the light field mixture Eqn.~\ref{eqn:mixture_field}, this is
\begin{equation}
  \begin{aligned}
      g_{\text{mix}}^{(2)}&(t_{2}-t_{1})=\\
     =  &\rho^{2}\,g_{\text{coh}}^{(2)}(t_{2}-t_{1}) + (1-\rho)^{2}\,g_{\text{unc}}^{(2)}(t_{2}-t_{1}) \\
    & + 2\rho(1-\rho) \left[1 + \Re[g_{\text{coh}}^{(1)}(t_{2}-t_{1})\,g_{\text{unc}}^{(1)*}(t_{2}-t_{1})\right]\,, \\
  \end{aligned}
  \label{eqn:g2_mixture}
% \end{split}
\end{equation}
where $g^{(1)}$ is the first-order field correlation function for the
respective component light fields, $g^{(1)*}$ its complex conjugate,
and $\Re[\cdots]$ extracts the real part of its argument.

The last term in Eqn.~\ref{eqn:expansion} can be written as
\begin{equation}
  \begin{aligned}
  &\langle {E_{\text{mix}}^{*}(t_{1})E_{\text{mix}}^{*}(t_{2}+\Delta)E_{\text{mix}}(t_{2})E_{\text{mix}}(t_{1}+\Delta)} \rangle \\
        =  &\rho^{2}\,\abs{g_{\text{coh}}^{(1)}(t_{2}-t_{1})}^{2} +
        (1-\rho)^{2}\abs{g_{\text{unc}}^{(1)}(t_{2}-t_{1})}^{2} \\
        + &2\rho(1-\rho)\,\Re[g_{\text{coh}}^{(1)}(t_{2}-t_{1})\,g_{\text{unc}}^{(1)*}(t_{2}-t_{1})] \\
        + &2\rho(1-\rho)\,\Re[g_{\text{coh}}^{(1)}(\Delta)\,g_{\text{unc}}^{(1)*}(\Delta)]\,,\\
  \end{aligned}
  \label{eqn:cross_terms_mixture}
\end{equation}
where $g^{(1)}(\Delta)\approx0$ 
for our experimental situation of the propagation delay $\Delta$ being significantly larger than the coherence times of the respective light sources.
Note that all terms in Eqn.~\ref{eqn:cross_terms_mixture} are
real-valued.  

With this, the interferometric photon correlation at
zero time difference in Eqn.~\ref{eqn:expansion} is given by
\begin{equation}
  \begin{aligned}
    g_{\text{mix}}^{(2X)}&(0)=\\
         = \frac{1}{4}[&g_{\text{mix}}^{(2)}(\Delta)+ g_{\text{mix}}^{(2)}(-\Delta)\\ 
        &+2(\rho^{2}\,g_{\text{coh}}^{(2)}(0) + (1-\rho)^{2}\,g_{\text{unc}}^{(2)}(0) + 2\rho(1-\rho)) \\
        &-2(\rho^{2}\,\abs{g_{\text{coh}}^{(1)}(0)}^{2} + (1-\rho)^{2}\abs{g_{\text{unc}}^{(1)}(0)}^{2})]\,.
  \end{aligned}
  \label{eqn:g2x_mixture_raw}
\end{equation}

We further assume that (1) the propagation delay in the interferometer $\Delta$ is significantly longer than the
coherence time scale of the light source, 
such that $g_{\text{mix}}^{(2)}(\pm\Delta)\approx1$,
(2) the interferometer has high visibility such that $\abs{g^{(1)}(0)}\approx1$,
and (3) the second order correlation of the coherent light field is
$g^{(2)}_\text{coh}(0)=1$. 
With this, Eqn.~\ref{eqn:g2x_mixture_raw}
leads to the relationship shown in Eqn.~\ref{eqn:g2x_mixture_extractable}.

%%%%%%%%%%%%%%%%%%%%%%%%
\section{Boundaries of physically meaningful combinations of interferometric
  correlations in a mixture}\label{app:boundaries}
%%%%%%%%%%%%%%%%%%%%%%%%
Assuming a binary mixture of the light field as per
Eqn.~\ref{eqn:mixture_field}, the interferometric correlation of the mixture,
$g_{\text{mix}}^{(2X)}(0)$, and the conventional second order correlation of
the incoherent light, $g_{\text{unc}}^{(2)}(0)$, at zero time difference are
constrained by relation Eqn.~\ref{eqn:g2x_mixture_extractable}. Further
assuming the physical requirement $0 \leq \rho \leq 1$ for the fraction $\rho$ gives a lower bound for $g_{\text{unc}}^{(2)}(0)$,
\begin{equation}
        g_{\text{unc}}^{(2)}(0) \geq  \begin{cases}0\,,  &g_{\text{mix}}^{(2X)}(0) \leq \frac{2}{3}\\
                        3 +\frac{1}{1 - 2 g_{\text{mix}}^{(2X)}(0) }, &g_{\text{mix}}^{(2X)}(0) \in[\frac{2}{3},1]  \\
                        2g_{\text{mix}}^{(2X)}(0)\, &g_{\text{mix}}^{(2X)}(0) \geq 1 \end{cases}\,.
\label{eqn:lowerbound_g2}
\end{equation}
For $g_{\text{mix}}^{(2X)}(0) \in [0,\frac{1}{2})$, there is an upper bound
\begin{equation}
    g_{\text{unc}}^{(2)}(0) \leq 2g_{\text{mix}}^{(2X)}(0)\,.
 \label{eqn:upperbound_g2}
\end{equation}
{}

%%%%%%%%%%%%%%%%%%%%%%%%
\section{Error propagation from fitting of $g^{(2X)}$ measurement}
%%%%%%%%%%%%%%%%%%%%%%%%
Standard error propagation techniques of experimental data through
Eqn.~\ref{eqn:fraction_upper_bound}-\ref{eqn:fit} lead to infinite
uncertainties for some dip amplitudes $A$ and are therefore not used.
Instead, we extract upper and lower bounds of $\rho$.
Equation~\ref{eqn:fraction_upper_bound}
provides an upper bound
\begin{equation}
     \rho \leq \sqrt{2A}\,,
 \label{eqn:fraction_upper_bound_A}
\end{equation}
and Eqn.~\ref{eqn:fraction_lower_bounds} the lower bound
\begin{equation}
    \rho \geq  \begin{cases}
    2A, &\text{for}\, 0 \leq A \leq \frac{1}{4}\\
    \frac{1}{2}+\frac{1}{2}\sqrt{4A-1}, &\text{for} \, \frac{1}{4} \leq A \leq \frac{1}{2}
    \end{cases}
 \label{eqn:fraction_lower_bounds_A}
\end{equation}
for $\rho$. The probability density for values of $A$ in a measured ensemble
is assumed to be a normal distribution, with a mean value and standard
deviation extracted from the fit of measured $g^{(2X)}$ to
Eqn.~\ref{eqn:fit}. This can be transformed into a probability distribution
for upper and lower bounds 
for $\rho$ using Eqns.~\ref{eqn:fraction_upper_bound_A} and
\ref{eqn:fraction_lower_bounds_A}. We exclude non-physical values of
$\rho$ outside $0 \leq \rho \leq 1$, and renormalize the resulting distribution
to compute an expectation value of $\rho$ and a $90\%$ confidence interval
shown in Fig.~\ref{fig:two_mode_vs_current}.

% \bibliography{references}
%apsrev4-2.bst 2019-01-14 (MD) hand-edited version of apsrev4-1.bst
%Control: key (0)
%Control: author (8) initials jnrlst
%Control: editor formatted (1) identically to author
%Control: production of article title (0) allowed
%Control: page (0) single
%Control: year (1) truncated
%Control: production of eprint (0) enabled
%

\end{document}